# ANALYSIS OF EFFICIENCY LIMITING PROCESSES IN THIN FILM Cu(In,Ga)(S,Se)$_2$ ELECTRODEPOSITED SOLAR CELLS

J.P.Connolly[1*], Z. Djebbour[2], A.Darga[2], C.Bazin[2], D. Mencaraglia[2], M. Benosman[1], N.Bodereau[1],
JF Guillemoles[1], D.Lincot[1], J. Kessler[1], N. Naghavi[1], J. Kurdi[1], O. Kerrec[1]
1 Institut de Recherche et Développement sur l'Énergie Photovoltaïque (IRDEP)
UMR 7174 CNRS-ENSCP-EDF 6 quai Watier 78401 Chatou
2 Laboratoire de Génie Électrique de Paris (UMR 8507, CNRS), LGEP-Supélec, Universités Paris VI &
Paris XI, Plateau de Moulon, 11 rue Joliot Curie, 91192 Gif-sur-Yvette, France
* Corresponding author: james-connolly@enscp.fr +33 (0) 1 30 87 76 56

ABSTRACT: Electrodeposited thin film cells have been fabricated with record-breaking efficiencies of 11.4%. This presentation examines conversion mechanisms in cells with a focus on the effect of CdS buffer layers using a range of complementary tools. Dark currents (IVs) are well described by series and parallel resistances, and two dominant recombination mechanisms represented by parallel diodes. Measurements of IV as a function of temperature (IVT) allow extraction of activation energies corresponding to these processes and indicate their spatial position. Admittance spectroscopy (AS) gives an independent estimate of the same energies, and yields values of the defect densities of states in the forbidden gap. Two dominant levels are apparent, confirming the validity of the IV analysis. Spectral response (QE) measurements are presented, yielding information on minority carrier collection efficiency. The different methods of parameter extraction are correlated and indicate recombination levels some hundreds of meV above the valence band and below the conduction band. Bias dependence of admittance spectroscopy gives indications on the localisation of defect centres with one defect situated at the CdS heterointerface and the other in the bulk of the depletion region. The dark current analysis indicates that photogenerated minority carrier collection is the limiting factor in these cells at the operating bias.

Keywords: Thin film, Modelling, CIGS

## 1 INTRODUCTION

Electrodeposited CIS thin film solar cells are a promising route to low cost solar energy with efficiencies above 11% [1]. The low cost aspect is maximised if vacuum deposition techniques are avoided. This is the case for the electrodeposition process [2] being elaborated by the CISEL project in the framework of the IRDEP institute.
In light of this relatively unexplored technology, it is important to re-evaluate loss mechanisms as a function of cell geometry. This work investigates cells with variable CdS buffer layer thicknesses in order to evaluate the dominant loss mechanisms.
We expect the analysis to confirm general design principles laid out in previous work. The analysis assumes that the cell structure is similar in all cases except for design differences intentionally introduced.
Characterisation tools include current voltage characteristics as a function of temperature (IVT), admittance spectra (CV), and relative quantum efficiency (QE). The IVT measurements are analysed in terms of a two-diode analysis with series and parallel resistance.

## 2 EXPERIMENTAL DETAILS

Samples consist of typical ZnO/CdS/CuInS2 structures. Samples were grown with CdS buffer layer thicknesses of 23nm, 40nm and 63nm, the window and absorber layers being fabricated in an identical manner. Details are given in ref. [3].

QE measurements were performed using a Jobin-Yvon monochromator, a tungsten 3000K light source and a Stanford lockin amplifier controlled by a control software in basic run on a 386 computer. IVT measurements were closed cycle cryostat and Oxford TC4 temperature controller. Admittance measurements were performed with an HP Agilent frequency analyser and an Agilent 42941.impedance spectrometer.

Samples were characterised with a class A Spectranova solar simulator in the light and in the dark.

## 3 ANALYSIS

### 3.1) Dark current fitting
Dark current fitting is done by an automated least squares variational minimisation fitting procedure.
The equivalent circuit expresses the dark current $I_D$ in terms of two diodes described by saturation currents $I_{o1}$, $I_{o2}$, ideality factors $n_1$ and $n_2$, and, series and parallel resistances $R_s$ and $R_p$ as follows.

$$I_D = I_{01}\left[\exp\left(\frac{q(V-I_DR_S)}{n_1 K_B T}\right)\right] + I_{02}\left[\exp\left(\frac{q(V-I_DR_S)}{n_2 K_B T}\right)\right] + \frac{(V-I_DR_S)}{R_P} \quad 1)$$

### 3.2) Admittance spectroscopy
Activation energies were estimated by applying the analysis of Walter *et al.* [4] developed by Nadenau *et al.* [5] for a band structure similar to our cells, with the interface between absorber and buffer producing a cliff type band line-up. Briefly summarised, dark currents are described in terms of tunnelling enhanced bulk and interface recombination. This is expressed in terms of a characteristic tunnelling energy Eoo, a characteristic temperature (or characteristic energy of the distribution) of defect state densities T*, and finally a tunnelling enhanced bulk recombination activation energy Ea.
The three samples were analysed by admittance spectroscopy yielding a spectrum of defect densities as a function of energy, and hence temperatures T*.
The dark current analysis of section 3.1) yields saturation current densities and ideality factors analysed along the lines of the theory outlined here, yielding activation energies along similar lines given by the slope of Arrhenius plots of ideality factor times saturation current versus inverse temperature.

### 3.3) Quantum efficiency
The quantum efficiency (QE) model SOLCIS calculates the QE for the ZnO-CdS-CIS p/n structure by solving the standard transport and continuity equations [6], shown

here for the minority electron population shown for completeness in the form

$$\frac{d^2 n_p}{dx^2} + \left(\frac{qE}{K_B T} + \frac{1}{D_n}\frac{dD_n}{dx}\right)\frac{dn_p}{dx} + \left(\frac{qE}{D_n K_B T}\frac{dD_n}{dx} + \frac{1}{K_B T}\frac{dE}{dx} - L_n^{-2}\right)n_p = -\frac{G(x)}{D_n} \quad 2)$$

Materials parameters in this system are variable and not well known due to the difficulty in isolating the different layers in order to measure relevant quantities. As a result the model is used as a comparative tool enabling qualitative conclusions to be drawn.

In this context, absorption coefficients and band structure are taken from available sources ([7], [8], [9], [10]) and diffusion lengths used as the single fitting parameter. The model is written in FORTRAN interfacing with PAW the freeware data analysis package designed at CERN. This model is work in progress with an aim to allow arbitrary position dependence of materials parameters and a two-dimensional modelling approach.

## 4    RESULTS AND DISCUSSION

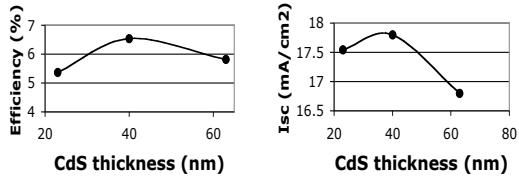

**Figure 1** Efficiency (left) and short circuit current under AM1.5 global illumination.

Figure 1 summarises cell performance and shows short circuit currents. Open circuit voltages and fill factors show similar trends and are not shown. The centre cell with a 40nm buffer shows the best performance.

4.1) Current voltage at different temperatures (IVT)

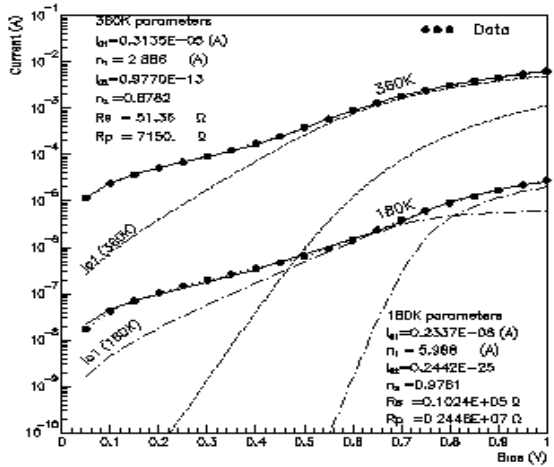

**Figure 2** 180K (left) and 360K (right) fits for 40nm CdS samples, showing data, fit, and two diode contributions

Good agreement is seen across the temperature range followed (figure 2). A different temperature activation for the two diode contributions is visible.

Figure 3 shows activation energies extracted from dark currents for all three thicknesses. The higher Shockley-Read-Hall [11] type ideality factor n1 appears at energy in the neighbourhood of the CdS barrier bandgap, indicating interface recombination dominating here. The second is as expected closer to the absorber bandgap and we relate this to bulk recombination assisted by tunnelling due to the large ideality factors seen.

Figure 4 shows the variation of the dominant saturation current Io1, essentially equivalent for the thinner CdS but lowest in the thickest CdS. This significantly indicates better interface quality with thicker CdS buffers,
Figure 5 shows the variation of dominant ideality factor n1 with temperature which as we will see in 4.2) closely agrees with the theory of ref. [5], discussed below.

Resistive behaviour shows similar exponential increase with inverse temperature in all cases as expected (not shown here). As expected, the parallel resistance is highest and therefore best for the thickest CdS sample. The 23nm and 40nm samples however show the same level of Rp of about 1E4 $\Omega$ for 0.1cm$^2$ total area samples. In this case this is not the efficiency limiting factor since the 40nm CdS cell is the most efficient.
The significant factor here is the reduced series resistance in the 40nm sample, which is discussed further below.

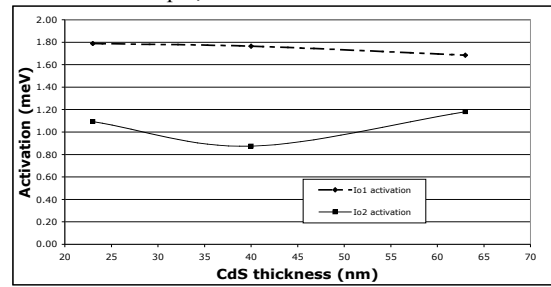

**Figure 3** Activation energies showing a decreasing Io1 saturation and a minimum in Io2 for the best cell.

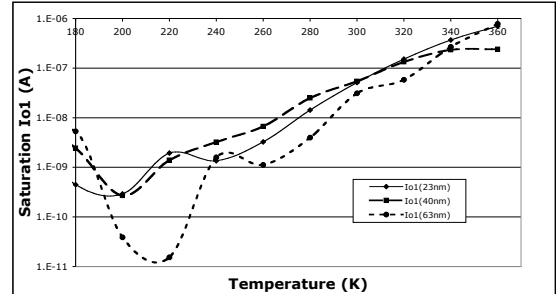

**Figure 4** Saturation currents Io1 for all temperatures

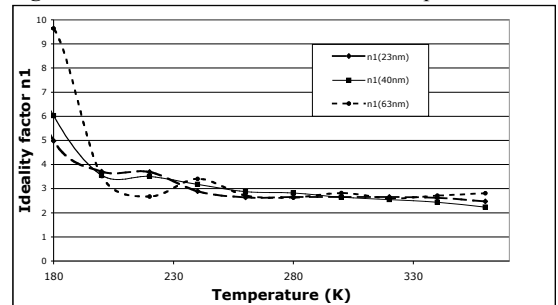

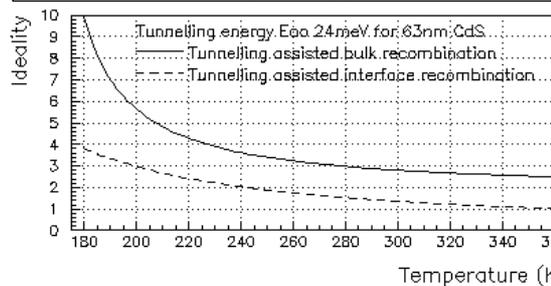

**Figure 5** Ideality factor n1 as a function of buffer thickness (top graph) and calculation for 63nm CdS [5].

| CdS (nm) | Defect type | Activation energy (meV) | Kb.T*/q (meV) | Peak DOS ($Cm^{-3}eV^{-1}$) |
|---|---|---|---|---|
| 25 | N1 | 93.01 | 14.97 | 1.75E+16 |
|    | N2 | 201.94 | 20.51 | 5.62E+16 |
| 40 | N1 | 105.75 | 17.28 | 2.08E+16 |
|    | N2 | 179.91 | 20.73 | 5.07E+16 |
| 63 | N1 | 101.95 | 18.65 | 1.00E+16 |
|    | N2 | 243.15 | 18.38 | 3.04E+16 |

**Table 1** Defect density distribution parameters

### 4.2) ADMITTANCE SPECTROSCOPY

The admittance spectra reveal two dominant defect densities. One appears situated in the bulk of the depletion by the width of the distribution, and the second at the CdS-CIS interface. The characteristic temperature T* (table 1) shows that defect N1 dominates the ideality factor of these samples and is attributed to a donor-like defect distribution near the conduction band edge and situated at the interface. This is correlated to the dominant dark current ideality factor extracted from IVT dark current fitting.

This is also shown in figure 5 which shows close agreement between ideality factors extracted from dark current fitting and ideality factors calculated following the formalism of ref. [4] and subsequent work [12]. The calculation uses a characteristic distribution energy (i.e. $K_BT^*/q$) of 20 meV throughout as in table 1. The single fitting parameter is the tunnelling energy.

This tunnelling energy is seen to vary slightly from 21 meV for 23 nm of CdS, to 24 meV for 63 nm.

The activation energies extracted here are not band-to-band energies presented in the dark current analysis but reflect activation energies from donor-like state distributions for N1 close to the conduction band Ec and acceptor-like distributions for N2 close to valence Ev [4]. Both are detected in the space charge region.

### 4.3) Quantum efficiency

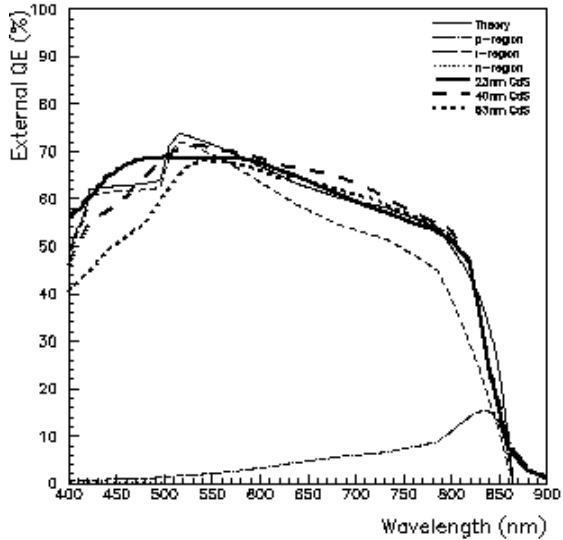

**Figure 6** Quantum efficiency for three CdS thicknesses and model for 23nm, showing the contribution from the neutral section of the absorber layer, the depletion "i" layer, and negligible ZnO and CdS contributions.

Figure 6 shows the measured QE for the three CdS thicknesses. The poor fit at short wavelength is due to imprecise ZnO and CdS absorption coefficients. The misfit is ascribed to an overestimation of the CdS absorption in particular above the CdS bandgap, and an underestimation of the absorption tail due to the spread expected in this nanocrystalline mixed phase material as opposed to more well defined values in the literature.

The first clear conclusion is that short wavelength QE decreases with increasing CdS thickness. The model predicts short circuit currents of 17.5, 17.3 and 17.1 mA/cm$^2$ versus 17.5, 17.8 and 16.8 mA/cm$^2$ from experiment for 23, 40 and 63nm CdS respectively under AM1.5G solar flux. This decrease is explained purely by absorption in the CdS which does not contribute to the photocurrent.

### 5) DISCUSSION

The three techniques presented approach the question of CdS buffer layer influence from different perspectives.

Dark current analysis gives a global overview of the recombination pathways We find two recombination processes similar to Shockley-Read-Hall and ideal Shockley [11] descriptions, with ideal Shockley increasingly dominant at low temperatures. With increasing CdS thickness, this dominant saturation decreases, indicating a more favourable depletion layer

Temperature dependence provides understanding in terms of models of tunnelling assisted recombination. Corresponding activation energies are related to interface recombination at the buffer - absorber interface, and absorber bandgaps respectively, identified from this perspective by the magnitude of the activation energies.

Ideality factors indicate significant tunnelling contributions to these recombination currents, and are well described by a formalism developed by ref. [5] illustrated in figure 5.

We see a slight reduction in dominant non interface recombination for the sample with the thickest CdS buffer layer, indicating that this is not the efficiency limiting factor here since the best cell is that with intermediate buffer thickness.

Parallel resistance similarly does not appear dominant since again the thickest buffer yields the highest parallel resistance and best or lowest leakage currents.

Series resistance however appears correlated to device performance with the best cell having the lowest series resistance. We have no reason however to explain why the median buffer thickness should have lowest series resistance, and conclude that this is incidental.

Admittance measurements yield defect densities as a function of energy, which allows extraction of defect activation energies in the depletion region. We find activation energies with similar trends to results reported in work using similar methods on other members of the CIGS materials family [12]. Here, we find the N1 conduction band donor-like defect activation is 100 meV, and the valence band acceptor-like level is 200 meV.

The defect density distribution is characterised by a characteristic energy expressed as a temperature T* which enables us to calculate ideality factors along the lines of the tunnelling assisted recombination formalism [5]. This defect density characteristic temperature is similar for all samples and allows estimation of the tunnelling energy. This is seen to be comparable for the 23 nm and 40 nm samples at 21 and 22 meV, and slightly higher for the thickest CdS buffer at 24 meV again indicating a slightly better junction due to slight increase in energy required for tunnelling.

Quantum efficiency analysis is straightforward. The quantum efficiencies indicate comparable minority carrier collection efficiency in the absorber. A first calculation of the quantum efficiency using available values of the absorption coefficients of CdS and ZnO describe the short wavelength QE loss well, confirming previous reports that the CdS buffer does not contribute to the photocurrent. From the point of view of short circuit current, therefore, the thinnest CdS buffer is the best.

Putting all this together, we see strong correlation between photocurrent and efficiency (fig. 1). We see also that increased buffer thickness improves the interface quality and reduces the dominant tunnelling assisted interface recombination.

CONCLUSIONS

The work presented here confirms standard understanding of the limiting dynamics of thin film CIS solar cells. That is, a sufficiently thick CdS buffer is necessary to prevent low parallel resistance, but that the buffer thickness must be minimised to prevent short circuit current loss.

We find close agreement between ideality factors as a function of temperature by two methods. First they are determined from experimental dark currents with a two diode model, and theoretical values from a model including tunnelling assisted recombination mechanisms.

IVT analysis shows dark currents are dominated at room temperature by tunnelling assisted recombination in the depletion region with an activation indicating location at the buffer-absorber interface. This agrees with admittance measurements which distinguish between defect close in energy to the conduction band and close to the CdS-CIS interface, and defect levels close to the valence band and more spatially distributed in the depleted bulk due the spread of the admittance spectra observed. The two methods agree but admittance provides information on spatial dependence in the depletion region which the IVT analysis does not.

A first look at efficiency parameters together with quantum efficiency variation and dark current analysis results, suggests that in these cells the efficiency is mainly limited by the short circuit current. At present we find an optimum at 40nm CdS, but this suggests that CdS deposition must be examined in detail to explain why the interface appears superior with greater buffer thickness, in order to design thinner CdS buffers with equivalent interface quality.

Future work envisaged is development of the QE model to take a two dimensional approach, and applying this analysis to the absorber layer. Furthermore, this analysis is worth repeating with a greater range of CdS thicknesses than reported here.

**Acknowledgements:**
We thank the European Union framework 6 program for a Marie Curie fellowship, making this work possible, and the ADEME for financial support for the CISEL project.